\documentclass[12pt,preprint]{aastex}

\slugcomment{Accepted for publication in AJ}

\shorttitle{ExploreNEOs VIII}
\shortauthors{Mommert et al.}

\begin{document}


\title{ExploreNEOs VIII: Dormant Short-Period Comets in the
  Near-Earth Asteroid Population}


\author{M. Mommert\altaffilmark{1,2}}

\author{A. W. Harris\altaffilmark{2}}

\author{M. Mueller\altaffilmark{3}}

\author{J. L. Hora\altaffilmark{4}}

\author{D. E. Trilling\altaffilmark{1,5,6}}

\author{W. F. Bottke\altaffilmark{7}}

\author{C. A. Thomas\altaffilmark{8}}

\author{M. Delbo\altaffilmark{9}}

\author{J. P. Emery\altaffilmark{10}}

\author{G. Fazio\altaffilmark{4}}

\and

\author{H. A. Smith\altaffilmark{4}}


\email{michael.mommert@nau.edu}

\altaffiltext{1}{Department of Physics and Astronomy,
Northern Arizona University, PO Box 6010, Flagstaff, AZ 86011}
\altaffiltext{2}{DLR Institute of Planetary Research,
  Rutherfordstrasse 2, 12489 Berlin, Germany}
\altaffiltext{3}{Netherlands Institute for Space Research, SRON,
  Postbus 800, 9700 AV Groningen, The Netherlands}
\altaffiltext{4}{Harvard-Smithsonian Center for Astrophysics, 60 Garden St., MS-65, Cambridge, MA 02138}
\altaffiltext{5}{Visiting Scientist, South African Astronomical Observatory}
\altaffiltext{6}{Visiting Professor, University of the Western Cape, South Africa}
\altaffiltext{7}{Southwest Research Institute,  1050 Walnut St, Suite
  300, Boulder, Colorado 80302, USA}
\altaffiltext{8}{ORAU/NASA Goddard Space Flight Center; 8800 Greenbelt Rd., Greenbelt, MD 20771}
\altaffiltext{9}{Universit\'e de Nice Sophia
Antipolis, CNRS, Observatoire de la C\^ote d'Azur, 06304 Nice
Cedex 4, France}
\altaffiltext{10}{Department of Earth and Planetary Sciences,
  University of Tennessee, 1412 Circle Dr.,
Knoxville, TN 37996}


\begin{abstract}
  We perform a search for dormant comets, asteroidal objects of
  cometary origin, in the near-Earth asteroid (NEA) population based
  on dynamical and physical considerations. Our study is based on
  albedos derived within the ExploreNEOs program and is extended by
  adding data from NEOWISE and the Akari asteroid catalog. We use a
  statistical approach to identify asteroids on orbits that resemble
  those of short-period near-Earth comets using the Tisserand
  parameter with respect to Jupiter, the aphelion distance, and the
  minimum orbital intersection distance with respect to Jupiter. From
  the sample of NEAs on comet-like orbits, we select those with a
  geometric albedo $p_V \leq 0.064$ as dormant comet candidates, and
  find that only ${\sim}$50\% of NEAs on comet-like orbits also have
  comet-like albedos. We identify a total of 23 NEAs from our sample
  that are likely to be dormant short-period near-Earth comets and,
  based on a de-biasing procedure applied to the cryogenic NEOWISE
  survey, estimate both magnitude-limited and size-limited fractions
  of the NEA population that are dormant short-period comets. We find
  that 0.3--3.3\% of the NEA population with $H \leq 21$, and
  $(9_{-5}^{+2})$\% of the population with diameters $d\geq 1$~km, are
  dormant short-period near-Earth comets.
\end{abstract}





\section{Introduction}

The population of near-Earth objects comprises small bodies, both
comets and asteroids, covering a wide range of dynamical parameters
and physical properties. This variety suggests that the members of the
population are a mixture of bodies of different origin and
evolution. The dynamical lifetime of near-Earth asteroids (NEAs),
which constitute the majority of the near-Earth object population, is
typically of the order of $10^7$~yrs \citep[e.g.,][]{morbidelli98},
which is significantly shorter than the age of the Solar System.
Therefore, the existence of the NEA population implies that there must
be sources of replenishment in order to maintain the observed
population. Source regions of NEAs have been identified to lie mostly
within the asteroid main belt and the transport mechanisms into the
NEA population are well understood \citep{wetherill79, wisdom83,
  vok00, bottke02}.

Comets have long been suspected of not only supplementing the cometary
component of the near-Earth object population, but also its asteroidal
component, the NEAs \citep{opik63, wetherill88, binzel92}. Comets are
objects from the outer regions of the Solar System that harbor ices
and have been perturbed by the gravitation of the giant planets into
orbits that bring them into the inner Solar System.  From the
dynamical viewpoint, there are two major populations of comets:
long-period comets with periods $P > 200$~yr and short-period comets
with periods $P \leq 20$~yr. Short-period comets have low inclinations
and interact strongly with Jupiter \citep{lowry08}; their
near-ecliptic orbits and short periods strongly suggest an origin in
or near the Kuiper belt, most probably in the scattered disk and
Centaur populations \citep{duncan04}. The orbits of long-period comets
are nearly isotropically distributed in inclination and have high
eccentricities, indicating an origin in the Oort cloud
\citep{lowry08}. Most Halley-type comets have periods $20 < P <
200$~yr and can be considered the short-period tail of the long-period
comets \citep{Weissman1996}. Their origin is still subject to debate;
models suggest an origin in the Kuiper Belt \citep{Levison2006} or the
Oort cloud \citep{Wang2014}.  In this work, we will focus on the
discussion of short-period comets in the near-Earth object population,
the short-period near-Earth comets (NECs).

As comets approach the Sun, the increased amount of insolation results
in a rise of their surface temperatures. Sublimation of near-surface
volatiles causes the development of cometary activity in the form of a
coma and a tail. \citet{levison97} found that the most likely activity
lifetime of short-period comets is ${\sim}12000$~yr, which is
significantly shorter than the average dynamical lifetime of
short-period comets \citep[$4.5\cdot10^7$~yr,][]{levison97} and NEAs
\citep[$10^7$~yr,][]{morbidelli98}.  Hence, comets that have spent a
significant amount of time in near-Earth space are likely to have
ceased their activity, becoming ``dormant'' or ``extinct'' comets that
are indistinguishable from low-albedo asteroids
\citep{wetherill91}. However, this is only one possible fate of
comets. Observations have shown that comets can break up into smaller
fragments \citep[see, e.g.,][]{Boehnhardt2004}, or, as recently
observed in comet C/2012~S1 (ISON), disrupt entirely. Results by
\citet{Whitman2006}, however, suggest that at the end of the active
lifetime of short-period comets they are likely to become dormant
rather than to disrupt.  \citet{levison97} estimate that 78\% of all
short-period comets are extinct. Additionally, there may be dormant
comets that presently appear asteroidal but could once again have a
cometary appearance. Examples include NEA 4015 Wilson-Harrington,
which displayed cometary activity in 1949, but never since
\citep{Bowell1992, Fernandez1997}, and NEA 3552 Don Quixote, which
was found to show cometary activity nearly 30 years after its
discovery as an asteroid \citep{Mommert2014}. Don Quixote has been
considered an extinct comet \citep{Hahn1985}, but it is more adequate
to describe it as a dormant comet, since it is not clear if its
activity is persistent and feeble or episodic. Accordingly, we adopt
the general term dormant comets, since it is not clear if all of these
objects are actually extinct.

Dormant comets in the NEA population, just like active comets, have
impacted Earth and are likely to have contributed to the deposition of
water and organic materials on its surface \citep[][and references
therein]{oro61,delsemme84,Mottl2007,Hartogh2011}. The determination of
the physical properties and the fraction of dormant comets in the NEA
population is important in order to understand the formation and
evolution of the Solar System and life on Earth. Comets are directly
linked to the outer regions of the Solar System, which contain the
most pristine objects. Since NEAs are among the most easily accessible
objects in space, dormant comets in near-Earth space provide us with
the unique opportunity to retrieve and study potentially volatile-rich
cometary material for future resource utilitzation.

In this work, we present a search for dormant comets that have an
origin as short-period comets based on a statistical analysis and an
estimation of their fraction in the NEA population. We base our
analysis on the largest sample of physically characterized NEAs
available to date.

\section{Identification of Dormant Comets}
\label{lbl:identification}

We identify dormant comet candidates in the NEA population using two
different statistical approaches that are based on the dynamical and
physical ensemble properties of known asteroids and comets. In our
first approach, we identify objects with comet-like orbits, utilizing
the Tisserand parameter with respect to Jupiter, $T_J$, and the
minimum orbit intersection distance with respect to Jupiter,
$MOID_J$. In our second approach we use $T_J$ and the aphelion
distance, $Q$, to identify objects on comet-like orbits. From both
samples, we then identify ``dormant comet candidates'' as objects with
low, comet-like albedos.

Our considerations are based on the sample of known near-Earth
asteroids and short-period near-Earth comets based on the JPL
Small-Body Database Search Engine\footnote{{\tt
    http://ssd.jpl.nasa.gov/sbdb\_query.cgi}} as of 28 April,
2015. Our sample includes 12533 NEAs and 65 NECs. Short-Period NECs
have been selected based on $q \leq 1.3$~au, $P < 20$~yr, and $2 \leq
T_J \leq 3$ (see Section \ref{lbl:tj} for a discussion); note that we
exclude comet fragments from our analysis.

\begin{figure}
 \centering
 \includegraphics[width=\linewidth]{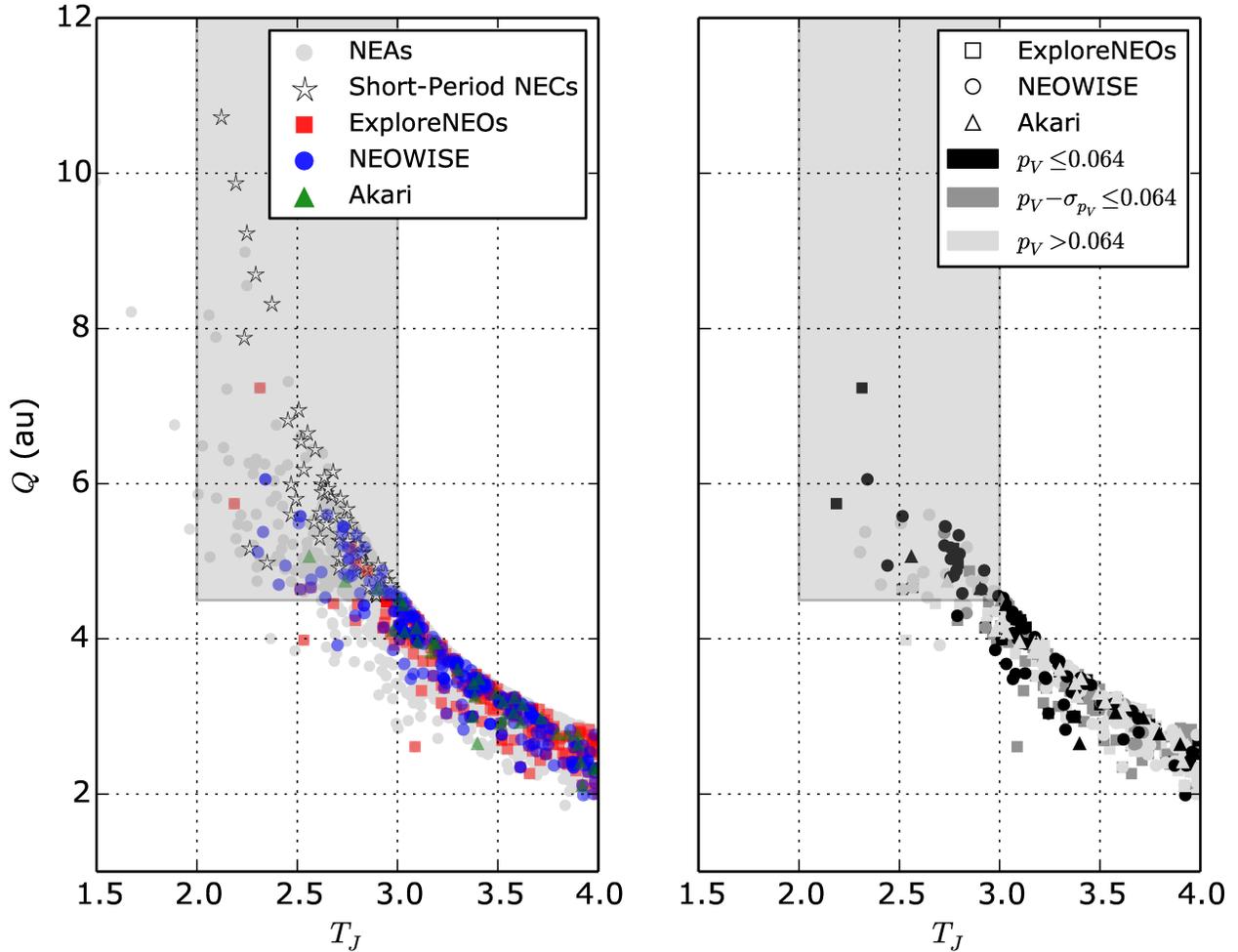}
 \caption{{\bf Left}: $Q$ as a function of $T_J$ for near-Earth
   asteroids and short-period near-Earth comets. Gray dots represent
   known NEAs, star symbols are known short-period NECs; red squares
   are ExploreNEOs targets, blue circles are NEOWISE targets, green
   triangles are Akari targets. We select dormant comet candidates
   from the shaded area, which contains all known short-period
   near-Earth comets with $2.0 \leq T_J \leq 3.0$ and 4.3\% of all
   known NEAs. {\bf Right}: Distribution of NEAs for which the albedo
   has been measured. Black symbols have albedos $p_V\leq0.064$, dark
   gray symbols agree with the albedo limit within one standard
   deviation, and light-gray symbols have higher albedos. We find 51\%
   of the NEAs with $Q \geq 4.5$~au and $2.0 \leq T_J \leq 3.0 $ to
   have comet-like albedos (see Section \ref{lbl:results}).}
 \label{fig:Q_tj}
\end{figure}

\begin{figure}
 \centering
 \includegraphics[width=\linewidth]{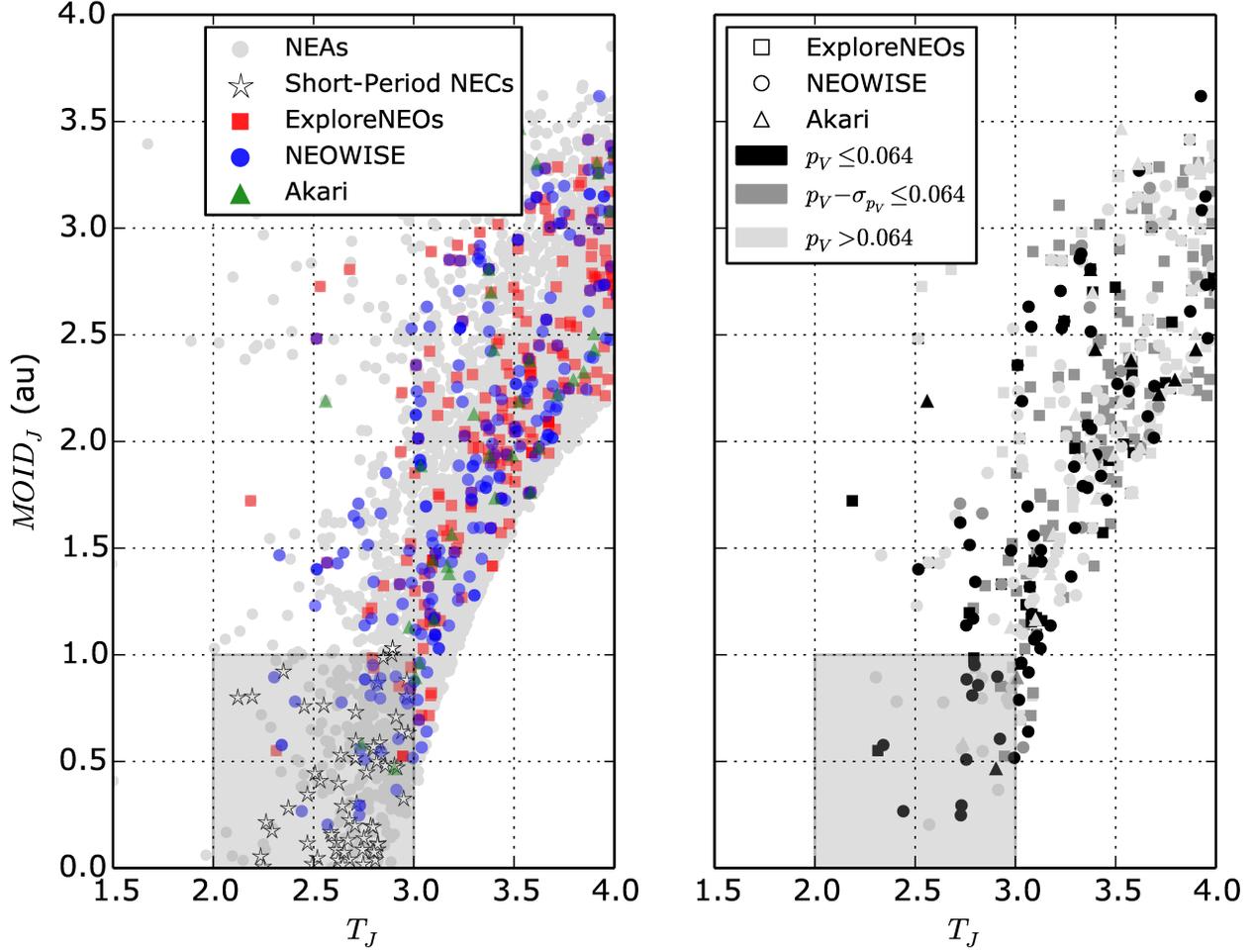}
 \caption{{\bf Left}: $MOID_J$ as a function of $T_J$ for near-Earth
   asteroids and short-period near-Earth comets. Gray dots represent
   known NEAs, star symbols are known short-period NECs; red squares
   are ExploreNEOs targets, blue circles are NEOWISE targets, green
   triangles are Akari targets.  We select dormant comet candidates
   from the shaded area, which contains 97\% of all known short-period
   NECs and 3.6\% of all known NEAs. {\bf Right}: Distribution of NEAs
   for which the albedo has been measured. Black symbols have albedos
   $p_V\leq0.064$, dark gray symbols agree with the albedo limit
   within one standard deviation, and light-gray symbols have higher
   albedos. In this case, we find 45\% of the NEAs with $2.0 \leq T_J
   \leq 3.0$ and $MOID_J \leq 1.0$~au to have comet-like albedos. }
 \label{fig:moid_tj}
\end{figure}


\subsection{Tisserand Parameter}
\label{lbl:tj}

The ``Tisserand parameter'' is a dynamical quantity that is diagnostic
of gravitational interaction of a body with a planet and is
approximately conserved during encounters in the restricted three-body
problem.
The Tisserand parameter with respect to Jupiter is defined as

\begin{equation}
  T_J = \frac{a_J}{a}+2\sqrt{\frac{a}{a_J}\left( 1-e^2
    \right)}\cos{i},
  \label{eqn:TJ}
\end{equation}

\noindent where $a$, $e$, and $i$ are the semimajor axis,
eccentricity, and inclination of the target body, respectively, and
$a_J$ is the semimajor axis of Jupiter \citep{Tisserand1896}. $T_J$ is
of special interest in orbital dynamics as it can be used as an
approximate discriminator between asteroidal ($T_J > 3.0$) and
cometary ($T_J \leq 3.0$) orbits. The actual $T_J$-boundary between
asteroids and comets is less strict, since $T_J$ is only conserved in
the idealized case of the restricted three-body problem. Comets with
$T_J > 3.0$ exist and are called ``Encke-type'' comets. Only a few
Encke-type comets are known and their origin is still subject to
debate \citep{levison05}. Halley-type and long-period comets usually
have $T_J < 2$. In this work, we will neglect both
long-period/Halley-type comets and Encke-type comets and only focus on
short-period NECs. \citet{levison97} define short-period comets as
comets with $2 < T_J < 3$, which allows the comets to experience
low-velocity encounters with Jupiter. Hence, such comets are
dynamically dominated by Jupiter, coining the term ``Jupiter family
comets''.

We use $T_J$ as determined by the JPL Small-Body Database Search
Engine for both NEAs and NECs.
For NEAs we use the same criterion that is used for the short-period
NECs: $2.0 \leq T_J \leq 3.0$. Previous works
\citep{fernandez05,demeo08,Kim2014} used the less strict criterion
$T_J \leq 3.0$, potentially including a number of Halley-type comets.

\subsection{Aphelion Distance}
\label{lbl:Q}

Comets have been scattered into the inner Solar System as a result of
close encounters with Jupiter \citep{levison97}. In order to be able
to have somewhat close encounters with Jupiter, any object is required
to have a sufficiently large aphelion distance $Q$ to feel the
gravitational pull of the giant planet. The distribution of comets in
$T_J$-$Q$ space (Figure \ref{fig:Q_tj}, left plot) suggests $Q \geq
4.5$~au, which we adopt as our criterion for a cometary orbit in
$Q$. Our criteria in $T_J$ and $Q$ apply to all short-period NECs but
only 4.3\% of the NEAs. Similar criteria have been adopted by
\citet{Kim2014}.

\subsection{Minimum Orbit Intersection Distance}
\label{lbl:moid}

The ``minimum orbit intersection distance'' with Jupiter, $MOID_J$,
describes the shortest distance between the orbit of a body and that
of the giant planet. Hence, it defines the distance of the closest
encounter both bodies can possibly have. \citet{sosa12} show that
comets in near-Earth space are more likely to have a low $MOID_J$ than
asteroids (their Figure 1). $MOID_J$ of both NEAs and short-period
NECs has been calculated using the code provided by
\citet{Wizniowski2013}. Since $MOID_J$ is not a dynamical invariant,
we can only look at a snapshot image of the dynamical characteristics
of the asteroid and comet populations. The deductions we can make are
still justified, since we use a statistical approach to identify
dormant comets in the NEA population. From Figure \ref{fig:moid_tj}
(left plot) it is obvious that most short-period NECs have $MOID_J
\leq 1.0$~au, which we adopt as our criterion. 97\% of the
short-period NEC population, and only 3.6\% of the known NEA
population meet this criterion.

\subsection{Albedo}

Cometary nuclei have low geometric albedos, $p_V$.
\citet{lamy04} compiled a list
of measured cometary nuclei albedos, most of which have $p_V \leq
0.05$. In their search for extinct cometary objects,
\citet{fernandez05} use an albedo upper limit of $p_R \leq 0.075$,
which is based on albedo determinations of comets in the $R$ band,
compiled in \citet{lamy04}, and an assumed albedo uncertainty of
30\%. We take an approach that is similar to that of
\citet{fernandez05}, and define an upper limit for cometary $V$-band
albedos based on previously measured albedos of short-period comets. In Table
\ref{tab:comet_albedos} we show the measured $V$-band albedos for the
small number of short-period comet nuclei for which this information has been
determined. From the measured albedos we determine the mean albedo
$\langle p_V \rangle$ to be $0.047\pm0.017$, where the uncertainty is
the quadratic sum of the standard deviation and the root mean square
of the uncertainties of the individual objects listed in Table
\ref{tab:comet_albedos}. Our approach to estimating this uncertainty
takes into account both internal and external uncertainties, i.e., it
includes the uncertainties of the individual albedo measurements as
well as the scatter of the ensemble of albedos. It is our intention to
determine an upper-limit albedo for short-period comets, so we define our albedo
limit as the mean value and add the uncertainty of 0.017, yielding
$p_V \leq 0.064$. This limit includes all the comet albedos listed in
Table \ref{tab:comet_albedos} and is comparable to, but slightly lower
than, the Fern\'{a}ndez value (we obtain $p_R = 0.071$ instead of
0.075), assuming a typical normalized spectral reflectivity gradient
for comets of $10\%/1000$\AA\ \citep{lamy04}.

\begin{deluxetable}{lcr}
\tabletypesize{\scriptsize}
\tablecaption{$V$-band Albedos of Short-Period Comets.\label{tab:comet_albedos}}
\tablewidth{0pt}
\tablehead{
\colhead{Name} & \colhead{$p_V$} & \colhead{Reference}}
\startdata
2P/Encke                &   0.046$\pm$0.023 &  \cite{fernandez00} \\
9P/Tempel~1             &   0.056$\pm$0.007 &  \cite{li07} \\
10P/Tempel~2            &   0.03$\pm$0.01   &  \cite{campins95} \\
22P/Kopff               &   0.042$\pm$0.006 &  \cite{lamy02} \\
28P/Neujmin~1           &   0.06$\pm$0.01  &  \cite{campins95} \\
49P/Arend-Rigaux        &   0.04$\pm$0.01  &  \cite{campins95} \\
67P/Churyumov-Gerasimenko & 0.059$\pm$0.02 &  \cite{Sierks2015}\\
81P/Wild~2              &   0.059$\pm$0.004 &  \cite{li09} \\
103P/Hartley~2          &   0.028$\pm$0.009 &  \cite{Lisse2009}\\
\enddata
\tablecomments{This compilation of $V$ band albedos yields an average value of $\langle p_V
\rangle = 0.047\pm0.017$. The albedo uncertainty of 49P has been recalculated based on
the diameter uncertainty given by \citet{campins95}.}
\end{deluxetable}

\section{NEA Sample}
\label{lbl:data_sample}

We base our search for dormant comets on NEA albedo measurements from
our Warm Spitzer ``ExploreNEOs'' program \citep{trilling10}, as well
as from the NEOWISE \citep{mainzer11c} and Akari \citep{Usui2011}
programs.

As part of ExploreNEOs we performed 589 observations
\citep[e.g.,][]{trilling10, mueller11, trilling15} of 562 different
optically discovered NEAs using the {\it Infrared Array Camera}
\citep{fazio04} onboard the {\it Spitzer Space Telescope}
\citep{werner04} at 3.6 and 4.5~$\mu$m. For each asteroid we derive
diameter and albedo using the Near-Earth Asteroid Thermal Model
\citep[NEATM,][]{harris98}. Three of our targets were not considered
in the papers quoted above due to detector saturation. 3552 Don
Quixote was saturated in both IRAC channels and displayed cometary
activity during the time of our observations \citep{Mommert2014}. Two
more targets, 4015 Wilson-Harrington and 52762 (1998 MT24) were
saturated in the 4.5~$\mu$m band, only. In order to derive flux
densities from saturated images, we fit a calibrated point-spread
function (PSF) model to the extended wings of the measured PSF,
ignoring the saturated parts of the image.  This method has been well
tested on a number of saturated observations of calibration stars
\citep[see, e.g.,][and references therein]{Mommert2014}. We add an
additional 5\% uncertainty in quadrature to those flux densities to
account for the increased calibration uncertainty. Three other
targets, 2004 QF1, 152952 (2000 GC2), and 162825 (2001 BO61) were too
faint to be detected in the 3.6~$\mu$m band, and results are based on
the 4.5~$\mu$m flux density measurement only.

In order to determine accurate albedos, precise measurements of the
absolute magnitude $H$ (the magnitude of an object at 1 au distance
from the observer and the Sun, and at zero phase angle) and of the
photometric slope parameter $G$ are crucial. In the ExploreNEOs
program, we obtain $H$ magnitudes from the JPL Horizons service
\citep{Giorgini1996}. The provided $H$ magnitudes are notoriously
unreliable \citep{juric02, pravec12} and do not come with uncertainty
estimates. Therefore, we replace these $H$ magnitudes by values taken
from peer-reviewed publications \citep{hagen12, pravec12} (87 updates
by Hagen, 29 by Pravec), where available. Where no measured values are
available, we have to rely on the JPL Horizons $H$ magnitudes.

We increase our sample size by adding albedo measurements from 471
observations of 409 different NEAs observed by the NEOWISE
program \citep{mainzer11c}, excluding those NEAs for which albedo
  has not been measured. The {\it Wide-field Infrared Survey
  Explorer} \citep[WISE,][]{Wright2010} carried out an all-sky survey
in its 3.4, 4.6, 12, and 22~$\mu$m bands. The design of the WISE
survey enables it to discover new NEAs and comets in the thermal infrared, which
minimizes the impact of albedo dependent discovery bias. NEOWISE also
measures the diameters and albedos of all detected objects using the
NEATM, using an approach that is similar to ExploreNEOs
  \citep{trilling10, mueller11, mainzer11c}. We use updated albedos of
  26 NEOWISE sample targets from \citet{pravec12} that are based on
  new measurements of $H$.

Furthermore, we add 59 NEAs from the ``Asteroid catalog using Akari''
\citep{Usui2011}, which is based on an all-sky survey in two
mid-infrared bands (S9W: 6.7-11.6~$\mu$m, L18W: 13.9-25.6~$\mu$m), from
which diameters and albedos have been derived using the Standard
Thermal Model \citep[STM,][]{Morrison1979,Lebofsky1986}. 

The combined samples comprise 1132 albedo measurements of 869
different NEAs, representing ${\sim}7\%$ of the known NEA population
as of April 2015. \citet{Usui2014} performed a comparison of
  diameters and albedos measured for main belt asteroids between IRAS,
  Akari, and NEOWISE. They find an average agreement within 10\% for diameter
  and 22\% for albedo measurements between the three surveys. A
  similar comparison of 110 NEAs observed by ExploreNEOs and NEOWISE
  shows an equally good agreement within 6\% in diameter and 22\% in
  albedo on average \citep{trilling15}.  

\section{Results}
\label{lbl:results}

Of the 869 different NEAs with measured albedo, 65 have $2.0 < T_J <
3.0$ (see discussion in Section \ref{lbl:data_sample}).  Of those, 43
meet our additional $Q$ criterion and 31 meet our $MOID_J$ criterion,
fulfilling our dynamical criteria for being cometary.  Their albedos
were analyzed in a second step: for both populations we performed a
weighted count of the number of objects with $p_V \leq 0.064$. Table
\ref{tbl:sample} lists all observations of asteroids for which at
least one albedo measurement (23 NEAs) with $p_V \leq 0.064$
exists. NEAs 385402~(2002~WZ2) and (2000~HD74) each have one
measurement with $p_V \leq 0.064$ and one with $p_V > 0.064$. This
discrepancy is most likely caused by lightcurve effects and the fact
that both measurements use the same $H$ magnitude, which is not
corrected for lightcurve effects. We reduce the weight of both objects
in the following analysis to 0.5, whereas all other objects for which
only $p_V \leq 0.064$ measurements exist have a weight of 1. For the
$Q$-selected objects we hence find a statistical weight of 22 for a
total of 43 NEAs with measured albedos; 22/43 = 51\% (of the sample
size) of the NEAs with $Q \geq 4.5$~au also have $p_V \leq 0.064$. For
the $MOID_J$-selected sample, the weighted count is 14/31 (45\% of the
sample size). Note that all $MOID_J$-selected asteroids with $p_V \leq
0.064$ are also in the sample of $Q$-selected asteroids with $p_V \leq
0.064$.

The overlap of those NEAs in the $Q$-selected (43 NEAs) and
$MOID_J$-selected (31 NEAs) samples with any albedo value is 24
objects, which is 56\% of the $Q$-selected and 77\% of the
$MOID_J$-selected sample. Figures \ref{fig:Q_tj} and \ref{fig:moid_tj}
(left plots) support the impression that in the $Q$-selected NEA
sample, the ratio of cometary over asteroidal objects might be lower
than for the $MOID_J$-selected sample. Nevertheless, the fact that in
both dynamically selected samples about 50\% of the objects have $p_V
\leq 0.064$ suggests a similar degree of mixing between potential
dormant comets ($p_V \leq 0.064$) and ordinary asteroids (any $p_V$)
for both dynamical criteria. We consider the $Q$-selected sample of
NEAs with $p_V \leq 0.064$, which includes the $MOID_J$-selected
sample in its entirety, to be our sample of dormant short-period comet
candidates in the NEA population.

\begin{deluxetable}{rccccccl}
\tabletypesize{\scriptsize}
\tablecaption{Dormant Short-Period Near-Earth Comet Candidates.\label{tbl:sample}}
\tablewidth{0pt}
\tablehead{
\colhead{Object Name} & \colhead{$d$} & \colhead{$p_V$} &
\colhead{$T_J$} & \colhead{$MOID_J$} & \colhead{$Q$} & \colhead{$H$} & 
\colhead{Source} \\
\colhead{} & \colhead{(km)} & \colhead{} &
\colhead{} & \colhead{(au)} & \colhead{(au)} & \colhead{(mag)} &
\colhead{} 
}
\startdata

3552 Don Quixote (1983 SA) & $18.4^{+0.3}_{-0.4}$ & $0.03^{+0.02}_{-0.01}$ & 2.314 & 0.551 & 7.234 & 13.0 & EN \\
5370 Taranis (1986 RA) & $5.31^{+0.08}_{-0.08}$ & $0.051^{+0.009}_{-0.009}$ & 2.731 & 0.294 & 5.446 & 15.2 & NW \\
5370 Taranis (1986 RA) & $6.3^{+0.5}_{-0.5}$ & $0.037^{+0.009}_{-0.009}$ & 2.731 & 0.294 & 5.446 & 15.2 & NW \\
20086 (1994 LW) & $4.8^{+1.3}_{-1.1}$ & $0.013^{+0.014}_{-0.007}$ & 2.770 & (1.197) & 5.168 & 16.9 & EN \\
248590 (2006 CS) & $4.7^{+0.8}_{-0.8}$ & $0.018^{+0.007}_{-0.007}$ & 2.441 & 0.267 & 4.945 & 16.6 & NW \\
385402 (2002 WZ2) & $2.3^{+0.7}_{-0.6}$ & $0.06^{+0.07}_{-0.03}$ & 2.515 & (2.482) & 4.638 & 17.0 & EN \\
385402 (2002 WZ2) & $1.6^{+0.1}_{-0.1}$ & ($0.11^{+0.03}_{-0.03}$) & 2.515 & (2.482) & 4.638 & 17.0 & NW \\
(2000 HD74) & $1.9^{+0.5}_{-0.5}$ & $0.03^{+0.03}_{-0.02}$ & 2.567 & (1.432) & 4.662 & 18.0 & EN \\
(2000 HD74) & $0.83^{+0.01}_{-0.01}$ & ($0.16^{+0.03}_{-0.03}$) & 2.567 & (1.432) & 4.662 & 18.0 & NW \\
(2001 HA4) & $1.85^{+0.04}_{-0.04}$ & $0.05^{+0.01}_{-0.01}$ & 2.772 & (1.515) & 4.814 & 17.6 & NW \\
(2004 EB) & $2.5^{+0.2}_{-0.2}$ & $0.04^{+0.01}_{-0.01}$ & 2.755 & (1.138) & 5.176 & 17.2 & NW \\
(2004 YR32) & $2.3^{+0.3}_{-0.3}$ & $0.031^{+0.007}_{-0.007}$ & 2.725 & (1.620) & 5.203 & 17.6 & NW \\
(2004 YZ23) & $9.4^{+4.3}_{-3.1}$ & $0.02^{+0.02}_{-0.01}$ & 2.186 & (1.722) & 5.742 & 15.2 & EN \\
(2009 KC3) & $2.2^{+0.5}_{-0.5}$ & $0.023^{+0.018}_{-0.018}$ & 2.728 & 0.248 & 5.451 & 18.0 & NW \\
(2009 WF104) & $2.23^{+0.03}_{-0.03}$ & $0.047^{+0.009}_{-0.009}$ & 2.800 & (1.342) & 5.096 & 17.2 & NW \\
(2009 WO6) & $2.49^{+0.01}_{-0.01}$ & $0.034^{+0.008}_{-0.008}$ & 2.785 & 0.810 & 4.881 & 17.3 & NW \\
(2009 XE11) & $2.72^{+0.02}_{-0.02}$ & $0.038^{+0.006}_{-0.006}$ & 2.796 & 0.952 & 5.336 & 17.0 & NW \\
(2010 AG79) & $0.89^{+0.01}_{-0.01}$ & $0.018^{+0.003}_{-0.003}$ & 2.814 & 0.858 & 4.587 & 20.2 & NW \\
(2010 DH77) & $0.63^{+0.02}_{-0.02}$ & $0.009^{+0.002}_{-0.002}$ & 2.516 & (1.401) & 5.581 & 21.8 & NW \\
(2010 DH77) & $0.52^{+0.02}_{-0.02}$ & $0.012^{+0.003}_{-0.003}$ & 2.516 & (1.401) & 5.581 & 21.8 & NW \\
(2010 FJ81) & $0.42^{+0.01}_{-0.01}$ & $0.049^{+0.009}_{-0.009}$ & 2.341 & 0.577 & 6.056 & 20.8 & NW \\
(2010 FJ81) & $0.5^{+0.1}_{-0.1}$ & $0.03^{+0.02}_{-0.02}$ & 2.341 & 0.577 & 6.056 & 20.8 & NW \\
(2010 FZ80) & $0.87^{+0.01}_{-0.01}$ & $0.018^{+0.004}_{-0.004}$ & 2.755 & 0.509 & 4.796 & 20.3 & NW \\
(2010 JL33) & $1.78^{+0.03}_{-0.03}$ & $0.047^{+0.009}_{-0.009}$ & 2.910 & 0.898 & 4.637 & 17.7 & NW \\
(2010 LR68) & $2.3^{+0.2}_{-0.2}$ & $0.017^{+0.004}_{-0.004}$ & 2.923 & 0.606 & 4.882 & 18.3 & NW \\
(2010 LV108) & $0.23^{+0.01}_{-0.01}$ & $0.029^{+0.005}_{-0.005}$ & 2.994 & 0.517 & 4.553 & 22.6 & NW \\
(2010 GX62) & $0.62^{+0.01}_{-0.01}$ & $0.041^{+0.007}_{-0.007}$ & 2.757 & 0.885 & 5.031 & 20.2 & NW \\
(2010 GX62) & $1.12^{+0.01}_{-0.01}$ & $0.012^{+0.002}_{-0.002}$ & 2.757 & 0.885 & 5.031 & 20.2 & NW \\
(2011 BX18) & $3.0^{+0.7}_{-0.7}$ & $0.012^{+0.011}_{-0.006}$ & 2.793 & 0.985 & 4.976 & 18.0 & EN \\

\enddata
\tablecomments{NEAs with orbits and albedos that resemble those of
  short-period NECs ($2 \leq T_J \leq 3.0$ and ($Q\geq4.5$~au or
  $MOID_J \leq 1.0$~au) and $p_V\leq 0.064$). For each object we list
  its diameter $d$, geometric albedo $p_V$, $T_J$, $MOID_J$, $Q$,
  absolute magnitude $H$, and source reference of the albedo
  measurement. Values in parentheses signal that the respective
  criterion has not been met (see Section \ref{lbl:identification}).}
\tablerefs{EN: ExploreNEOs \citep[this
  work;][]{trilling10,trilling15,Mommert2014}; NW: NEOWISE
  \citep{mainzer11c,pravec12}; AK: Akari \citep{Usui2011}}
\end{deluxetable}

\begin{figure}[t!]
 \centering
 \includegraphics[width=\linewidth]{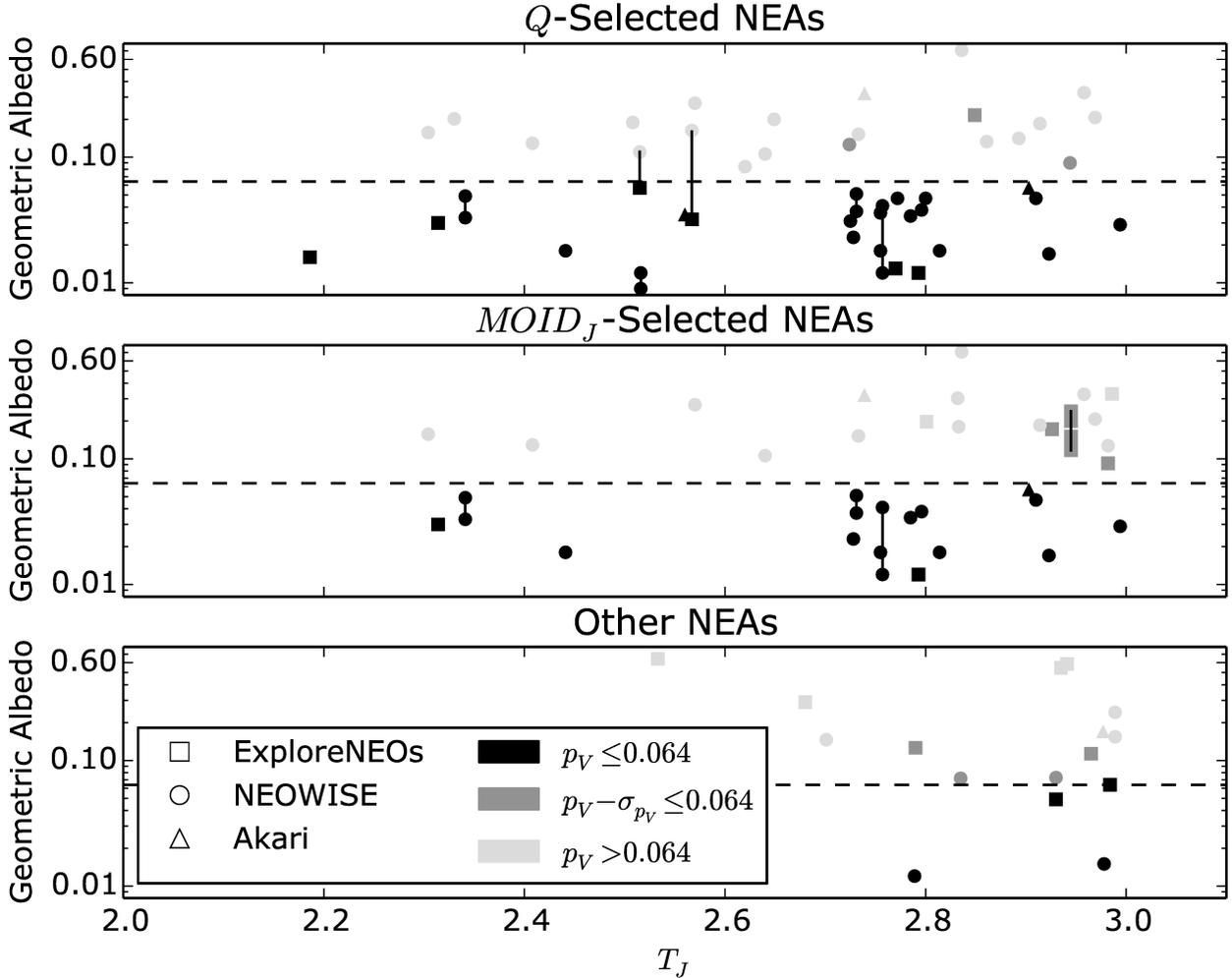}
 \caption{Albedos of NEAs with $2.0 \leq T_J \leq 3.0$ as a
   function of $T_J$ for samples selected based on $Q$ and $MOID_J$,
   as well as the exclusive sample (``Other NEAs''). Black symbols
   refer to NEAs with $p_V \leq 0.064$, dark gray symbols to
   objects with $p_V - \sigma \leq 0.064$ (where $\sigma$ is the lower
   $1\sigma$ albedo uncertainty), and light gray symbols to objects
   with $p_V > 0.064$. The dashed line indicates $p_V = 0.064$,
     our albedo upper limit for comet-like albedos. For objects with
   more than one albedo measurement, the individual datapoints are
   connected with lines. 
   We do not find any low-albedo NEAs with $T_J \leq 2.8$ that
   are neither in the $Q$ nor the $MOID_J$-selected samples. For the
   sake of readability, error bars are not shown; typical albedo
   relative uncertainties are of the order of 50\%.}
 \label{fig:pv_tj}
\end{figure}

Figure \ref{fig:pv_tj} compares the albedo distributions of different
samples of NEAs with $2.0 \leq T_J \leq 3.0$ as a function of
$T_J$. In both the $Q$ and $MOID_J$-selected samples there is no
obvious trend of lower albedo with decreasing $T_J$, since high
albedos can be found irrespective of $T_J$. 
Interestingly, we do not find any NEAs with $p_V \leq 0.064$ for $2.0
\leq T_J \leq 2.8$ that is not in either the $Q$ or the
$MOID_J$-selected samples (Figure \ref{fig:pv_tj}). We discuss the
implications of the different albedo distributions in Section
\ref{lbl:high_albedos}.

\section{Discussion}
\label{lbl:discussion}

\subsection{Assessment of the Dormant Comet Fraction in the NEA Population}
\label{lbl:cometary_content}

Based on our identification of dormant short-period near-Earth
comet candidates in Section \ref{lbl:results}, we investigate the
fraction of dormant short-period NECs in the NEA
population. The discovery of dormant comets through optical surveys,
which discover the majority of NEAs, is hampered for two
reasons: highly eccentric, comet-like orbits mean that dormant comets
spend most of their time far from the Earth, and their low albedos
limit their optical brightness. Both factors have to be taken into
account to obtain a reliable estimate of the dormant comet content.

In order to minimize the impact of discovery bias, we base this
analysis of the dormant comet fraction solely on the NEOWISE
sample. The nature of the NEOWISE survey as an all-sky survey in the
thermal infrared provides a uniform sample of the NEA population that
is much less affected by albedo bias than optical surveys
\citep[e.g.,][]{mainzer11c}. Using technical details on the NEOWISE
survey from \citet{Wright2010} and \citet{mainzer11c}, we produce a
NEOWISE survey simulator in order to de-bias the NEA population as
observed by WISE. For a detailed description of the survey simulator
we refer to Appendix \ref{lbl:simulator}.

Using our NEOWISE simulator, we derive the dormant short-period NEC
fraction in a magnitude-limited and a size-limited sample of the NEA
population. Table \ref{tbl:sample} lists only a few dormant comet
candidates with $H>21$ that were observed by NEOWISE. Hence, we decide
to restrict our magnitude-limited sample to $H\leq21$, which includes
17 dormant comet candidates (see Table \ref{tbl:sample}) with
diameters larger than ${\sim}400$m, assuming an albedo of 0.047. In
order to properly account for albedo and absolute magnitude
uncertainties of each object, we vary both parameters according to
Gaussian statistics. We adopt the measured albedo uncertainty (see
Table \ref{tbl:sample}) as the 1$\sigma$ uncertainty, and in the case
of the absolute magnitude, we adopt a 1$\sigma$ uncertainty of
0.2~mag, which is based on results by \citet{juric02}. From 100 trials
with varied physical properties, we find that 0.3--3.3\% (1$\sigma$
confidence interval) of the NEA population with $H\leq21$ can be
considered dormant short-period NECs.

For our size-limited estimate of the dormant short-period NEC fraction
we consider NEAs with diameters $d \geq 1$~km, a size range that is
well-sampled by NEOWISE and nearly entirely discovered
\citep{mainzer11c}. Using the same method as above, we vary the
diameter and albedo within the NEOWISE-derived uncertainties and run
the simulation 100 times. We find that $(9_{-5}^{+2})$\% (1$\sigma$)
of the NEA population with $d\geq1$~km are dormant short-period
NECs. In combination with the estimate of the number of NEAs with $d
\geq 1$~km by \citet{mainzer11c}, we conclude that ${\sim}$100 NEAs
with diameters of 1~km or more are dormant comets.

We note that these estimates differ significantly. We attribute this
discrepancy to two different effects. Due to the albedo-dependence of
the absolute magnitude and the wide range of albedos in the NEA
population, high-albedo NEAs are over-represented in the
magnitude-selected sample. Hence, one would expect a smaller fraction
of dormant comets in the magnitude-selected sample, compared to the
size-selected one. Furthermore, one has to account for the different
slopes of the size-frequency distributions of NEAs and comets
\citep[compare to, e.g.,][]{Meech2004, mainzer11c, trilling15}, the latter
of which is more shallow due to the disintegration of small cometary
objects with low perihelion distances. We find that the uncertainties
we obtain for both fractions are well within the confidence ranges we
derive in Appendix \ref{lbl:simulator_consistency_check}.

We compare our findings with earlier estimates of the dormant comet
fraction in the NEA population. \citet{bottke02} used dynamical
simulations of a de-biased synthetic NEA population to estimate the
fraction of cometary objects in the NEA population and found a
fractional content of ($6\pm4$)\% in the magnitude-limited NEA
population with $13 < H < 22$. Our magnitude-limited estimate,
0.3--3.3\%, is lower than but partly overlaps with their estimate,
covering nearly the same range in magnitude (we use $H\leq21$).
\citet{fernandez05} based their assessment on albedo measurements of
10 dynamically selected NEAs. Their sample selection was solely based
on observability, is therefore not de-biased, and has to be assumed to
be magnitude-limited. From their target sample they selected objects
on comet-like orbits with $T_J\leq3.0$ and albedos $p_R
\leq 0.075$. They find 4\% of all NEAs to be of cometary origin. Our
magnitude-limited result, 0.3--3.3\%, is slightly lower than their
result. \citet{demeo08} based their analysis on 39 NEAs with $T_J \leq
3.0$. In order to identify cometary object candidates, they either
require $p_R \leq 0.075$ or a C, D, T, or P taxonomic classification
of the object. Furthermore, they base their result on the assumption
that 30\% of all NEAs with $d\geq1$~km have $T_J \leq 3.0$
\citep{stuart03}. They find that ($8\pm5$)\% of all NEAs with
$d\geq1$~km are dormant comets. We find that $(9_{-5}^{+2})$\% of NEAs
with $d\geq1$~km are dormant comets which is in good agreement with
their result. \citet{Whitman2006} estimate a total of ${\sim}75$
dormant comets in the NEA population with $H < 18$. Assuming an albedo
of 0.047, this magnitude limit is equal to sizes of $d > 1.5$~km. Our
size-limited estimate infers a total of ${\sim}100$ dormant comets in
the NEA population with $d\geq1$~km, which is of the same order of
magnitude.

\subsection{Albedo Distribution of NEAs on Comet-Like Orbits} 
\label{lbl:high_albedos}

In Figure \ref{fig:pv_tj} we compare the albedo distributions of the
$Q$ and $MOID_J$-selected samples as a function of $T_J$. 
\citet{fernandez05} found that of the 9 NEAs they analyzed with $2.0
\leq T_J \leq 3.0$ (6 of their own from which we exclude 2008~OG108
for being a comet, and 4 from the literature), 44\% have nominal
albedos $p_R \leq 0.075$ and 66\% have albedos $p_R \leq 0.075$ within
the uncertainties. We find that ${\sim}$50\% of NEAs on comet-like
orbits also have comet-like albedos, which is in good agreement with
their result. \citet{fernandez05} have 2 NEAs in their sample with
$T_J \leq 2.6$, both of which have comet-like albedos, potentially
suggesting that NEAs with $T_J \leq 2.6$ are more likely to have
comet-like albedos.
We find that the ratio of comet-like albedos to
  non-comet-like albedos is approximately constant for the intervals
  $2.0 \leq T_J \leq 2.6$ and $2.6 \leq T_J \leq 3.0$. Hence, the findings
from this work and others \citep{fernandez05, Kim2014} suggest that
the albedo distribution of NEAs on comet-like orbits is less strictly
correlated to the dynamical distribution than previously
expected. This heterogeneity implies that not all NEAs on comet-like
orbits have a cometary origin. Potential asteroids that move on
supposedly comet-like orbits have been identified by
\citet{Fernandez2014}, who performed orbital integrations of a sample
of NEAs on comet-like orbits. Most of these objects, which probably
originate from the asteroid main belt, are likely to have higher than
cometary albedos, accounting for the observed albedo diversity.

We also compare the albedo distributions of NEAs in the $Q$ and
$MOID_J$-selected samples with those ``other'' NEAs that are in
neither of the two samples. We find that none of the ``other'' NEAs
with measured physical properties and $2.0 \leq T_J\leq 2.8$ has a low
albedo ($p_V\leq0.064$). We estimate the probability of any
interlopers, non-cometary NEAs with $2.0 \leq T_J\leq 2.8$ and
$p_V\leq0.064$, to not meet either the $Q$ or the $MOID_J$
criterion. Based on the 23 dormant short-period NEC candidates with
$2.0 \leq T_J \leq 2.8$, the probability of a newly discovered NEA
with the same properties not to be dormant short-period NEC candidate
is ${\leq}1/(23+1) = 4\%$. We conclude that any NEA with $2.0 \leq
T_J \leq 2.8$ and $p_V\leq0.064$ has a ${\geq}$96\% probability to be
of cometary origin.

\section{Conclusions}

From our search for dormant short-period near-Earth comets in the NEA population we can draw
the following conclusions:

\begin{itemize}
\item We identify 23 NEAs with orbits and albedos resembling
    those of short-period NECs that can be considered dormant
    short-period NECs.
\item From a de-biasing of the NEOWISE survey, we find that 
    0.3--3.3\% of the NEAs with $H\leq21$ and $(9_{-5}^{+2})$\% of
  those with $d\geq1$~km can be considered dormant short-period
    NEC candidates. The magnitude-limited fraction is slightly
  lower than earlier estimates, whereas the size-limited fraction
  agrees with earlier estimates. We estimate that ${\sim}$100
    NEAs with diameters of 1~km or more are dormant short-period
    NECs.
\item 
  We find that only ${\sim}$50\% of our sample NEAs on short-period
  NEC-like orbits have comet-like albedos, suggesting mixing between
  cometary and asteroidal objects among our sample targets. However,
  we find that any NEA with $2.0 \leq T_J \leq 2.8$ and $p_V\leq0.064$
  has a ${\geq}$96\% probability to be of cometary origin.
\end{itemize}

\acknowledgments

M.\ Mommert acknowledges support by the DFG Special Priority Program
1385, ``The First 10 Million Years of the Solar System - a Planetary
Materials Approach''. We would like to thank two anonymous
  referees for useful suggestions that led to significant
improvements of the manuscript. Support for this work was
  provided by NASA awards NNX10AB23G and NNX12AR54G. This work is
based in part on observations made with the Spitzer Space Telescope,
which is operated by the Jet Propulsion Laboratory, California
Institute of Technology under a contract with NASA. This publication
makes use of data products from the Wide-field Infrared Survey
Explorer, which is a joint project of the University of California,
Los Angeles, and the Jet Propulsion Laboratory/California Institute of
Technology, funded by the National Aeronautics and Space
Administration. This research is based on observations with Akari, a
JAXA project with the participation of ESA.
Support for this work was provided by NASA through an award
issued by JPL/Caltech.



{\it Facilities:} \facility{Spitzer(IRAC)}.

\appendix

\section{NEOWISE Survey Simulator}
\label{lbl:simulator}

We simulate the detectability of NEAs through the WISE all-sky survey
\citep{Wright2010,mainzer11c} in order to account for biases inherent
to the survey. The results of the simulator are utilized to obtain a
picture of the NEA population that is much more complete and less
prone to discovery bias. We determine the detection efficiency for WISE
as a function of orbital and physical properties of the NEA, based on
a simulated input NEA population, a simplified model of the WISE
observation strategy \citep{Wright2010}, and the WISE detection
efficiency in its most sensitive bands, W3 and W4
\citep{mainzer11c}. The de-biased NEA population is then derived by
dividing the sample of NEAs that were actually observed by WISE in its
cryogenic mission phase \citep{mainzer11c} by the derived
detection efficiencies. We base this analysis solely on the cryogenic
part of the WISE mission, which provides data from the most sensitive
bands (W3 and W4).

\subsection{Method}

Each object of the input NEA population is characterized by a set of
orbital parameters (semimajor axis, $a$, eccentricity, $e$,
inclination, $i$, the longitude of the ascending node, $\Omega$, the
argument of the perihelion, $\omega$, and the mean anomaly, $M$, at
the epoch) and physical properties (absolute magnitude, $H$, albedo,
$p_V$, diameter, $d$, and thermal model beaming parameter $\eta$). Our
input NEA population consists of 100000 NEAs derived with the NEA
model by \citet{Greenstreet2012}, to each of which we randomly assign
physical properties that are in accordance with the distributions in
$H$, $p_V$, and $\eta$ found by \citet{mainzer11c}. The synthetic
input NEA population is summed up in a matrix $S = (a \times e \times
i \times d \times p_V)$, according to their orbital and physical
properties, i.e., each cell of the matrix holds the number of objects
with a specific set of properties. Note that $d$ can be replaced by
$H$ in matrix $S$ if the simulation is performed on a
magnitude-limited instead of a size-limited sample; for the sake of
simplicity we will only use $d$ in the following discussion. The set of
actual NEOWISE detections throughout the cryogenic part of the WISE
mission is read into a similar matrix, $R = (a \times e \times
i \times d \times p_V)$.

For each object in the input NEA population we determine its
geocentric position for each day during the cryogenic part of the WISE
mission (January 14 -- August 5, 2010; 203~days) using the Python
package PyEphem\footnote{{\tt
    https://pypi.python.org/pypi/ephem/}}. WISE orbits in a low-Earth
polar orbit and observes at 90\degr\ solar elongation
\citep{Wright2010}. Hence, we determine times for which each object is
in quadrature relative to the Earth. For each quadrature situation, we
perform a more thorough check for WISE observability: for each 11~sec
timestep \citep[WISE observation cadence,][]{Wright2010}, we check if
the object's position coincides within 23.5\arcmin\ (half-width of the
WISE field of view) of the WISE pointing. The declination of WISE is
assumed to follow a polar rotation with a period of 94.3~min, which
has been derived from the orbital properties given by
\citet{Wright2010}. In accordance to the NEOWISE moving object
pipeline specifications \citep{mainzer11c}, we require each
potentially detectable NEA to appear in at least 5 fields and to have
a moving rate of 0.06--3.2\degr\ day$^{-1}$. This approach is
simplified in such a way as it neglects the details of WISE pointing,
e.g., with respect to the Moon.

In a second step, we estimate the thermal-infrared brightness of each
potentially detectable object during each observability window.
For each occassion in which an object is present in the WISE field of
view, we derive its thermal emission at wavelengths 11.5608~$\mu$m
(W3) and 22.0883~$\mu$m (W4) using the Near-Earth Asteroid Thermal
Model \citep[NEATM,][]{harris98} and based on the physical and orbital
properties of the object. The predicted thermal flux densities are
compared to the measured sensitivity of the respective band
\citep[Equation 3 in][]{mainzer11c} and the detection probability in
each band ($P_{W3}$, $P_{W4}$) is derived. The final detection
probability of one object is defined as $max(P_{W3}, P_{W4})$
over the cryogenic mission phase.

The detection probabilities of all objects from the input NEA
population are summed up in a matrix similar to $S$, $P = (a \times e
\times i \times d \times p_V$). The
detection efficiency is derived as $E =
P/S$ in an element-wise matrix division. Finally, the
de-biased population, $D$, is derived by dividing the sample of NEAs
observed during the cryogenic part of the WISE mission by the
efficiency matrix, $D=R/E$ (element-wise).

\subsection{Consistency Check}
\label{lbl:simulator_consistency_check}

We test the consistency of the NEOWISE simulator with the real NEOWISE
survey using two different tests. The first test compares the
compatibility of simulator detections with the real survey using two
exclusive samples: the sample of objects that was detected during the
cryogenic part of the NEOWISE program \citep{mainzer11c} (``NEOWISE
sample'', 471 detections) and those objects that were observed by the
{\it Spitzer Space Telescope} in the framework of the ExploreNEOs
program \citep{trilling10}, but not by the NEOWISE program
(``ExploreNEOs-not-NEOWISE'', 460 detections). Ideally, the NEOWISE
simulator detects all objects in the NEOWISE sample and none of those
objects in the ExploreNEOs-not-NEOWISE sample. For this test, we use
the actually measured physical properties (diameter, albedo, and
$\eta$), as well as the real orbital elements for each object and run
the simulation over the duration of the cryogenic WISE mission
phase. We derive the completeness of each sample by summing up the
detection probabilities of the individual objects and divide the sum
by the total number of objects in each sample. We find that 88\%
of all NEOWISE-observed NEAs (12\% false negatives) are detected
by our simulator and only 7\% of the ExploreNEOs-not-NEOWISE objects
(7\% false positives). Hence, the overall agreement in NEA
detectability between the simulator and the real NEOWISE survey is
good.

In a second test, we investigate the accuracy of our de-biasing
technique by replicating the estimate of the number of 1~km-sized NEAs
derived by \citet{mainzer11c}. We perform 100
simulator trials in which we vary the physical properties of the NEAs
listed in \citet{mainzer11c} within their uncertainties based on
Gaussian statistics. In the case of objects that have more than one
detection, we randomly reject duplicate detections such that every NEA
appears only once in matrix $R$. We
de-bias the 1~km NEA population by comparing the number of such
objects that were detected by NEOWISE ($R_{d\geq1\mathrm{~km}}$) with the number of objects in
our synthetic input population ($S_{d\geq1\mathrm{~km}}$). We find an
average number of 1~km-sized NEAs of 1200${\pm}$150, which agrees
within 1.5$\sigma$ with the number found by \citet{mainzer11c},
981$\pm$19. Note that \citet{mainzer11c} based their result on a
more elaborate de-biasing technique, which takes into account more
information about the known NEA population, whereas our approach can
be considered a ``blind'' de-biasing that is only based on those NEAs
detected by the NEOWISE survey. Hence, we consider the significance of
the agreement between these results adequate.

\end{document}